\documentstyle[preprint,aps]{revtex}%
\tighten
\begin{document}
\draft
\preprint{\vbox{Submitted to Phys.~Rev.~Lett. \hfill FSU-SCRI-95-37 \\
                                       \null\hfill IU/NTC 95-05  \\
                                       \null\hfill nucl-th/9505001}}
\title{Charge-Symmetry-Breaking Potentials from \\
       Isospin-Violating Meson-Baryon Coupling Constants}
\author{S. Gardner and C.~J. Horowitz}
\address{Nuclear Theory Center and Department of Physics, \\
         Indiana University, Bloomington, IN 47505}
\author{J. Piekarewicz}
\address{Supercomputer Computations Research Institute, \\
         Florida State University, Tallahassee, FL 32306}
\date{\today}
\maketitle

\begin{abstract}
We consider charge-symmetry violations in the nucleon-nucleon
force which result from isospin-violating meson-baryon coupling
constants. The vector mesons are assumed to
couple to the nucleon's electromagnetic
current, which we decompose into isoscalar and isovector quark
components. We compute these currents
in the context of a constituent
quark model. The isospin violations
in the meson-baryon couplings arise from the difference in the up and down
constituent quark masses. We show that class IV
charge-symmetry-breaking potentials arise in the resulting
$\omega$ and $\rho$ exchange contributions to the $NN$ force. The
magnitude of these contributions is consistent with that
phenomenologically required by the
measured difference of $n$ and $p$ analyzing powers
in elastic $\vec{n}-\vec{p}$ scattering at 183 MeV.
\end{abstract}

\narrowtext

Isospin violation manifests itself in a number of hadronic and nuclear
observables. The
scattering length differences of the $pp$ and $nn$ systems~\cite{slaus89},
the binding energy differences of mirror nuclei~\cite{miller90,nolsch69},
and the
$n$ and $p$ analyzing power difference in elastic $\vec{n}-\vec{p}$
scattering~\cite{knut90,abegg89,abegg94}
are all examples.
These effects presumably originate
in the differing mass and electromagnetic interactions of the
up and down quarks.
Consequently,
the confrontation of theoretical calculations of isospin-violating
observables
with experiment is of continuing interest, as it potentially
grants us new insight into
hadronic structure and
offers constraints on phenomenological models of
QCD.

Our focus will be on
charge-symmetry-breaking (CSB) in the $np$ system, which is
generated by the so-called class IV
CSB potentials~\cite{henmil79}. In contrast, the class III CSB
potentials respect isospin in the $np$ system, but distinguish $pp$ from
$nn$ systems.
There have been several calculations of
these potentials in the context of meson exchange
models~\cite{miller86,holz87,willia87}.
In such a picture, three distinct CSB contributions
to the $NN$ force exist.
CSB contributions can arise from (i) isovector-isoscalar
mixing in the meson propagators, (ii) isospin-breaking in the
meson-nucleon coupling constants, and (iii) isospin-breaking in
the nucleon wave function. In addition, there are electromagnetic
contributions, such as the photon's coupling
to the neutron's anomalous magnetic moment.
In principle, all these effects
contribute to CSB observables; one wishes to combine them
in a dynamical model.
Several different sources of CSB
contribute to the non-zero analyzing power difference
$\Delta A\equiv A_n - A_p$ measured
in polarized, elastic $n-p$ scattering.
Those studied so far include: the exchange of charged pions and rhos, the
photon's coupling to the neutron's anomalous magnetic moment, and
$\rho - \omega$ mixing.
The last is large because of the small mass difference between the $\rho$ and
$\omega$; the exchanged $\rho$ can convert into an $\omega$.
This mixing is clearly seen in $e^+e^- \rightarrow \pi^+\pi^-$
cross section measurements at the $\omega$ production point~\cite{barkov85}.
The $\rho-\omega$ mixing amplitude which fits the $e^+e^-$ data also
explains the $\Delta A$ measurement at 183 MeV~\cite{knut90} and accounts
for a large fraction of the binding energy difference seen
in the $A=3$ systems~\cite{miller90}.
However, it has been suggested that the $\rho-\omega$ mixing
amplitude depends on the momentum transfer $q$~\cite{ght92}.
Indeed, several authors
argue that the $q^2$ dependence is large and that the resulting
isospin-violating potential is
small at the space-like momentum transfers relevant for CSB
experiments~\cite{piewil93,hats93,krein93,mitch94,oconn94,maltman95}.
The issue continues
to be controversial~\cite{miller94,oconn95,urech95}.

If the $\rho-\omega$ mixing potential is, in fact,  small, then
the CSB contributions discussed so far no longer suffice to fit the
data. Yet other sources of isospin violation could well exist. Indeed, one
ought to consider isospin violation arising
from the nucleon's intrinsic wave function
as well as from the vector-meson-nucleon coupling constants.
These are
sources of additional isospin violation; they deserve
examination regardless of the $q^2$ dependence of the $\rho-\omega$
mixing amplitude.

In this paper we focus on isospin violation in the
vector meson couplings to the
nucleon.
As most CSB studies have focused on the role of mechanisms
(i), that is, $\rho-\omega$ mixing, and (iii)
--- through the sensitivity of charged pion and rho
exchange to the nucleon mass difference --- discussed above,
we study (ii) exclusively, as we wish to understand
its impact.
We assume that isospin violation in the nucleon's
internal wave function,
while undoubtedly nonzero, is negligibly small
due to the large mass difference
between the nucleon and the $\Delta(1910)$ ---
the first $P_{31}$ baryon.
The $\rho-\omega$ mass difference, in contrast, is a mere 12 MeV.
There is no argument, however, which protects the isospin symmetry
of the vector-meson-nucleon coupling constants.
In the following we examine the isospin violation arising from
the mass difference of the up and down quarks.
Electromagnetic radiative corrections have been estimated
earlier~\cite{yakin79}.

Dmitra\v{s}inovi\'c and Pollock
have studied the isospin-violating electroweak form factors of the
nucleon in a simple constituent quark model~\cite{dmitra95}.
These are potentially
important for interpreting parity-violating electron-nucleon scattering
in terms of the nucleon's strange
quark content, as the $Z^0$\ coupling is sensitive to isospin
violation.  They find that the isoscalar quark current $\bar{u} \gamma_\mu u
+ \bar{d} \gamma_\mu d$ has a larger matrix element in the proton than in
the
neutron as the up quark has a larger magnetic moment than a down quark.
The size of the violation,
which is about one percent, is set by the ratio of the difference in
up and down constituent quark masses to
their average mass.

Henley and Zhang have calculated the isospin dependence of the
vector-meson-nucleon couplings
in a constituent quark model, through explicit calculation of
the quark model wave function overlaps~\cite{henzha87}.
Our results for the
isospin-violating couplings are
very similar to theirs.

  Two assumptions define our model. First, the vector mesons are
assumed to couple to the appropriate isospin components of the
nucleon's electromagnetic current. This assumption is in the
spirit of the vector meson dominance model. Second, we assume this
current can be estimated --- at low $q^2$ --- in a nonrelativistic,
constituent quark
model. Such models give good descriptions of the nucleon magnetic moments.

  Perhaps the simplest way to realize these assumptions is in a
hybrid quark-meson model, in which the mesons couple directly to the
quarks. However, this picture is not required. A model with composite
vector mesons can still satisfy our assumptions.

In our model the vector mesons couple to the nucleon's electromagnetic
current, which we decompose into isoscalar and isovector quark
components, appropriate for the coupling of the $\omega$ and $\rho$,
respectively, to
the nucleon.
In the quark model,
the isoscalar electromagnetic charge of the up and
down quarks is $e_i^{(0)}=1/3$, whereas
the isovector electromagnetic charge of the up quark
is $e_u^{(1)}=1$ and that of the down quark is $e_d^{(1)}=-1$.
The vector quark current is
\begin{equation}
  J^{\mu} = e_{u}\, \bar{u} \gamma^{\mu} u +
            e_{d}\, \bar{d} \gamma^{\mu} d \;;
 \label{vcurrent}
\end{equation}
the constituent quarks are assumed elementary.
We are interested in computing the vector coupling $g^V_N$
and the tensor coupling $f^V_N$
of the nucleon to
the vector mesons $\rho$ and $\omega$.
That is,
\begin{eqnarray}
  \langle &N(p',s')& |J_V^{\mu}(q) | N(p,s) \rangle = \nonumber \\
  &\bar{U}(p',s')& \left[
      g^V_N \gamma^{\mu} +
    i f^V_N \sigma^{\mu\nu} {(p'-p)_{\nu} \over 2M_N}
  \right] U(p,s) \;.
 \label{formf}
\end{eqnarray}
Note that $U(p,s)$ denotes an on-shell nucleon spinor of
mass $M_N$, momentum $p$,
and spin $s$.
The couplings
$g^V_N$ and $f^V_N$ are functions of the four-momenta at the
vertex, here
$g^V_N(q^2)$ and $f^V_N(q^2)\; (q\equiv p'-p)$, though
we presume the couplings constant in our region of interest.
We compute the couplings at $q^2=0$, as
the nonrelativistic quark model is best-suited to an estimate
in the static limit.
For low-energy scattering experiments, such as
the 183 MeV $\vec{n}-\vec{p}$ analyzing power measurement~\cite{knut90},
this limit should be reasonable.
We obtain $g^V_N$ and $f^V_N$ by examining the nonrelativistic
reduction of Eq.~[\ref{formf}] and then computing the matrix elements of the
resulting operators in the quark model.
Thus, we evaluate
\begin{mathletters}
\label{themodel}
 \begin{eqnarray}
{   g^V_N \over  \overline{g}^V }
&=& \sum_{i=1}^3
    \langle N \!\uparrow | e^{(\tau)}_i | N\! \uparrow\rangle
 = \sum_{i=1}^3 e^{(\tau)}_i \\
   {(g^V_N + f^V_N )\over {2M_N
           \bar{g}^V }} &=&
   \sum_{i=1}^3
   \langle N \!\uparrow | {e^{(\tau)}_i \over 2m_i} \sigma^z_i|
    N\! \uparrow\rangle
 \end{eqnarray}
\end{mathletters}
in the nucleon rest frame.
One sums over the charges and magnetic moments of the quark $i$.
The symbol $e^{(\tau)}_i$ denotes the appropriate isospin component
of the electromagnetic charge of the quarks; the
$\omega$-nucleon coupling, for example, is determined by the isoscalar
quark charge.
The couplings $g^V_N$ and $f^V_N$ are written explicitly in units of
           $\overline{g}^V$, the isospin-averaged vector coupling of the
vector mesons to the nucleon.
Note that
           $\overline{g}^{\omega}$
and
           $\overline{g}^{\rho}$ are
known from fits to $NN$ scattering and to the
properties of the deuteron~\cite{machl87,machl89}. The
ket $| N\! \uparrow\rangle $ denotes a nucleon state with spin up.
We are interested in evaluating the isospin-violating contributions
to $g^V_N$ and $f^V_N$ and use the full $SU(6)$
wave function for the nucleon~\cite{close79}.
In this limit, the magnetic
moments are independent of the spatial distribution of the wave function,
so that they follow immediately from the spin structure
of the nucleon.
Note that $g^V_N$, in contrast, depends
only on the nucleon's flavor structure.
Consequently, the calculation of the vector and tensor couplings
proceeds straightforwardly.
The difference between the up and down quark masses can generate isospin
violations in the meson-baryon couplings. Introducing
\begin{equation}
   m \equiv {1 \over 2}(m_{d}+m_{u}) \;; \quad
   \Delta m \equiv (m_{d}-m_{u}) \;,
\end{equation}
equation (\ref{themodel}) implies that
\begin{equation}
g_N^{\omega} = \overline{g}^{\omega} \quad ; \quad
g_N^{\rho} = \overline{g}^{\rho}
\end{equation}
and, defining
\begin{mathletters}
\label{fresult}
\begin{eqnarray}
{f_N^{V} \over 2M_N}
&\equiv& {{f^V_{(0)}  + f^V_{(1)}\tau_z} \over 2M } \;, \\
\hbox{that} \qquad\qquad\qquad\qquad\qquad\quad
&\quad& \nonumber \\
f^{\omega}_{(0)} = 0 \quad &;& \quad
f^{\omega}_{(1)} = { 5\over 6} {\Delta m \over m}
\overline{g}^{\omega} \\
f^{\rho}_{(0)} = { 3\over 2} {\Delta m \over m}
\overline{g}^{\rho}
 \quad &;& \quad
f^{\rho}_{(1)} = 4\overline{g}^{\rho} \;,
\end{eqnarray}
\end{mathletters}
where $M=(M_n + M_p)/2$ denotes the mean nucleon mass --- the
isospin breaking we compute
includes the effect of
the neutron-proton mass difference.
We have chosen $m=M/3=313$ MeV in
Eqs.~(\ref{fresult}b) and (\ref{fresult}c).
We adopt this choice throughout the
paper. Note that $\tau_z$ acts at the hadronic level, so that
$\tau_z\;|p\rangle = +|p\rangle$ and so on.
The isospin-breaking corrections contribute to the tensor
couplings exclusively --- the vector couplings are unchanged.
Moreover, these corrections are {\it isovector}
for the $\omega$ coupling, and are {\it isoscalar} for the $\rho$
coupling. Thus, their appearance simulates $\rho-\omega$ mixing; this
will become explicit when we discuss the resulting
CSB potentials.
Note that $\Delta m > 0$ in the constituent
quark model~\cite{licht89}; the up quark, which is lighter, generates
a larger anomalous magnetic moment for the proton.

   Before discussing the isospin-breaking corrections in detail, let
us consider the isospin-symmetric results for $g_N^V$ and $f_N^V$.
That is,
\begin{equation}
{f^{\omega}_{(0)} \over g^{\omega}_N} =0 \;\; ; \;\;
{f^{\rho}_{(1)} \over g^{\rho}_N} =4 \;.
\end{equation}
These nonrelativistic quark model (NRQM)
results are qualitatively consistent with the $f^V_N/g^V_N$
ratios which emerge from phenomenological fits to the $NN$
interaction~\cite{machl87,machl89} --- recall that the Bonn B
potential parameters~\cite{machl89}, for example, are
$f^{\omega}_N/g^{\omega}_N=0$  and $f^{\rho}_N/g^{\rho}_N=6.1$.
These successes are intimately connected to the NRQM's ability to
describe the nucleon magnetic moments. In the above model,
the anomalous magnetic moment is purely isovector:
$\kappa_N = 2 \tau_z$. Note that
$\kappa_p^{\rm exp}=1.79$ and $\kappa_n^{\rm exp}=-1.91$.
The above successes encourage us to use the NRQM to compute
the isospin-violating corrections to these coupling constants as well.
These corrections are given in Eq.~(\ref{fresult}).

  Henley and Zhang have also examined the impact of quark mass
difference effects on the vector-meson-nucleon coupling
constants~\cite{henzha87}. They adopt an ``effective perturbative
QCD model'': they calculate the nucleon-nucleon-meson vertex
in terms of nonrelativistic, constituent quarks and connect the
produced $q\overline{q}$ pair with the other quarks via perturbative
one-gluon exchange.
We are able to reproduce the isospin breaking they compute in the
vector-meson-nucleon coupling constants.
The isospin breaking of their model can apparently be generated on rather
general grounds.

  We shall now compute the CSB potentials which arise from the
isospin-violating couplings
in Eq.~(\ref{fresult}). In a one boson exchange approximation,
we obtain the following CSB potentials for $\omega$ and $\rho$ exchange:
\begin{mathletters}
\label{classiv}
\begin{eqnarray}
V_{\rm CSB}^{\omega} &=&
- \left(
{ f_{(1)}^{\omega} \overline{g}^{\omega}
\over q^2 - m_{\omega}^2 }
\right)
{\hat{\cal{V}}}(1,2)
\;, \\
V_{\rm CSB}^{\rho} &=&
- \left(
{ f_{(0)}^{\rho} \overline{g}^{\rho}
\over q^2 - m_{\rho}^2 }
\right)
{\hat{\cal{V}}\;\!'}(1,2) \;,
\end{eqnarray}
\end{mathletters}
where
${\hat{\cal{V}}}(1,2)=
 \Gamma^{\mu}(1) \gamma_{\mu}(2) \tau_z(1) -
 \gamma^{\mu}(1) \Gamma_{\mu}(2) \tau_z(2)$ and
$\Gamma^{\mu} = i \sigma^{\mu \nu} q_{\nu} / 2 M$ with
$q=(p'_1 -p_1)$. Note that
${\hat{\cal{V}}\;\!'}(1,2)$ is of the form of
${\hat{\cal{V}}}(1,2)$ with the exchange
$\tau_z(1)\leftrightarrow \tau_z(2)$.
The above are identical in form to the CSB potential from $\rho-\omega$
mixing. That is,
\begin{equation}
\label{vrhoom}
V_{\rm CSB}^{\rho-\omega} =
-
{\overline{g}^{\omega}
f_{(1)}^{\rho}
\langle \rho | H | \omega \rangle
\over (q^2 - m_{\rho}^2 )(q^2 - m_{\omega}^2 )}
{\hat{\cal{V}}}(1,2) \;.
\end{equation}
%
%where $\overline{f}^{\rho}$ denotes the isospin-averaged
%$\rho$-nucleon tensor coupling.
In the case of $\rho$ exchange,
the couplings of Eq.~(\ref{fresult})
also generate a class III CSB potential.
The CSB potentials of Eq.~(\ref{classiv}) can be
combined in the nonrelativistic limit
to yield the class IV potential~\cite{henmil79}
\begin{eqnarray}
\label{vrhoomsim}
&V_{IV}^{\rho + \omega}&(q^2 \rightarrow 0) =
\left({5 \over 6}
{{\overline{g}^{\omega}}^2 \over m_{\omega}^2}
-{3 \over 2}
{{\overline{g}^{\rho}}^2 \over m_{\rho}^2}
\right)
\left( {\Delta m \over m }\right)
\nonumber \\
&\times&
{i\left(\vec{\sigma}(1) - \vec{\sigma}(2)\right)
\cdot \vec{q}\times\vec{P} \over 4M^2}
\left(\tau_z(1) - \tau_z(2)\right) \\
&\equiv& C (q^2=0)
{i\left(\vec{\sigma}(1) - \vec{\sigma}(2)\right)
\cdot \vec{q}\times\vec{P} \over 4M^2}
\left(\tau_z(1) - \tau_z(2)\right)
\nonumber
\;,
\end{eqnarray}
where $\vec{P}=\vec{p}_1\;\!' + \vec{p}_1$.
Let us compare the strength of the CSB potentials given in
Eqs.~(\ref{vrhoom}) and (\ref{vrhoomsim}).
The $C(q^2=0)$ of Eq.~(\ref{vrhoomsim}),
in terms of the Bonn B
potential parameters ($\overline{g}_{\omega}^2(q^2=0)/4\pi=11.13$;
$\overline{g}_{\rho}^2(q^2=0)/4\pi=.42$)~\cite{machl89}, is
\begin{eqnarray}
\label{geffres}
C(q^2=0)&=& 1.77 \cdot 10^2
{\Delta m \over m} \;{\rm GeV}^{-2} \nonumber \\
& & \\
&\approx& 2.32\;{\rm GeV}^{-2} \nonumber \;.
\end{eqnarray}
Note that the sign of Eq.~(\ref{geffres}) is determined by the
$\omega$ contribution --- in the Bonn model
${\overline{g}^{\omega}}^2 / {\overline{g}^{\rho}}^2 \approx 27$.
The last estimate for $C$ results when one uses the ``lower bound'' of
$\Delta m$, $\Delta m=4.1$ MeV, of Lichtenberg~\cite{licht89}.
The strength of the class IV $\rho-\omega$ mixing
potential in Eq.~(\ref{vrhoom}) at $q^2=0$, on the other hand,
is
\begin{eqnarray}
\label{rhoom}
C_{\rho-\omega}^{\rm on-shell} (q^2=0) &=&
-{\overline{g}^{\omega} f_{(1)}^{\rho}
\over m_{\rho}^2 m_{\omega}^2}
\langle \rho | H | \omega \rangle \vert_{q^2=m_{\omega}^2}
\nonumber \\
& & \\
&\approx&2.07\;{\rm GeV}^{-2} \nonumber \;,
\end{eqnarray}
where we have used the {\it on-shell} value of the $\rho-\omega$
mixing matrix element,
$\langle \rho | H | \omega \rangle = -4520 \pm 600\;
{\rm MeV}^2$~\cite{barkov85},
for purposes of comparison.  Note that we have also used the
Bonn B value for $f_{(1)}^{\rho}$,  $f_{(1)}^{\rho}=6.1\overline{g}^{\rho}$.
A potential of the magnitude of Eq.~(\ref{rhoom})
is needed for a successful
description of the 183 MeV $\Delta A$ data~\cite{knut90}.
Thus, isospin violation
in the meson-baryon coupling constants suffices {\it alone} to
generate the qualitative magnitude of
the phenomenologically required class IV CSB potential.

Here we have focused on the class IV CSB potential which arises from
isospin violations in the vector-meson-nucleon coupling constants.
The resulting class IV CSB potential
is identical in structure to that which arises
from $\rho-\omega$ mixing. Moreover, its magnitude
is commensurate in size with that phenomenologically required to
explain the IUCF $\Delta A$ measurement~\cite{knut90}.
If the $\rho-\omega$ mixing amplitude is $q^2$ dependent and the
isospin-violating potential
small for the space-like momentum transfers relevant to the above
experiment~\cite{ght92,piewil93,hats93,krein93,mitch94,oconn94,maltman95},
then we have found a source of isospin violation
which can fill the role demanded by the data. If the $\rho-\omega$
mixing amplitude is not $q^2$ dependent~\cite{miller94}, then the
total CSB potential is probably too large to
fit the data.

  The isospin-breaking we compute in the vector-meson-nucleon
coupling constants at $q^2=0$ arises on rather general grounds.
We assume that the vector-mesons couple to the appropriate isospin
components of the electromagnetic current; we compute these
components of the current in the nonrelativistic constituent quark
model. The magnitude of the isospin-breaking we predict
depends numerically
on only $\overline{g}_{\omega}$
and $\Delta m/m$.
The results we obtain do not
depend on the details of the nucleon's structure in the NRQM. Indeed,
our results depend merely on the manifest
spin and flavor structure of the nucleon
in the $SU(6)$ limit. Consequently, we believe our estimate to
have little model-dependence. This is why we reproduce the
isospin-breaking of Henley and Zhang's more complicated quark
model~\cite{henzha87}.
The isospin-breaking we predict could
have a nontrivial $q^2$ dependence. This requires
a detailed model calculation beyond the scope of our present
approach.

\acknowledgments

We thank V.~Dmitra\v{s}inovi\'c and S.J.~Pollock for sharing their
work prior to publication and for many important discussions, crucial
to the work presented here. J.P. acknowledges S.~Capstick for helpful
conversations. This work was supported by the DOE under
Contracts Nos. DE-FG02-87ER40365 (S.G. and C.J.H.),
DE-FC05-85ER250000 (J.P.), and DE-FG05-92ER40750 (J.P.).

\end{document}